\documentclass[12pt,aps,prd,preprint,preprintnumbers,tightenlines,
superscriptaddress,showpacs]{revtex4}
\newcommand{\PRE}[1]{{#1}} 

\usepackage{bm} 
\usepackage{epsfig}

\newcommand{\kev}{\text{keV}}
\newcommand{\mev}{\text{MeV}}
\newcommand{\gev}{\text{GeV}}
\newcommand{\tev}{\text{TeV}}

\newcommand{\eqref}[1]{Eq.~(\ref{#1})}

\newcommand{\br}{\tilde{\pi}}
\newcommand{\bl}{\tilde{\pi}^{\text{l}}}
\newcommand{\bh}{\tilde{\pi}^{\text{h}}}

\begin{document}



\preprint{UCI-TR-2007-53}

\preprint{FTPI-Minn-07/37}

\preprint{UMN-TH-2628/07}

\title{
\PRE{\vspace*{1.5in}}
Diffuse MeV Gamma-rays and Galactic 511 keV Line \\ from Decaying WIMP Dark Matter 
\\
\PRE{\vspace*{0.3in}} }

\author{Jose~A.~R.~Cembranos}
\affiliation{
William I. Fine Theoretical Physics Institute,
University of Minnesota, Minneapolis, 55455, USA
\PRE{\vspace*{.5in}} }
\author{Louis E.~Strigari%
\PRE{\vspace*{.2in}} } 
\affiliation{Department of Physics and
Astronomy, University of California, Irvine, CA 92697, USA
\PRE{\vspace*{.5in}} }
\begin{abstract}
\PRE{\vspace*{0.3in}} 

The origin of both the diffuse high-latitude MeV gamma-ray emission 
and the 511 keV line flux from the Galactic bulge are uncertain. Previous 
studies have invoked dark matter physics to independently explain these
observations, though as yet none has been able to explain both of these 
emissions within the well-motivated framework of Weakly-Interacting Massive 
Particles (WIMPs). Here we use an unstable WIMP dark matter model to show 
that it is in fact possible to simultaneously reconcile both of these observations, 
and in the process show a remarkable coincidence: decaying dark matter with 
MeV mass splittings can explain both observations if positrons and photons are 
produced with similar branching fractions. We illustrate this idea with an unstable 
branon, which is a standard WIMP dark matter candidate appearing in brane world 
models with large extra dimensions. We show that because branons decay via three-body 
final states, they are additionally unconstrained by searches for Galactic MeV 
gamma-ray lines. As a result, such unstable long-lifetime dark matter particles provide 
novel and distinct signatures that can be tested by future observations of MeV gamma-rays.

\end{abstract}


\maketitle

\section{Introduction}

The existence of dark matter is well-established, yet its identity remains elusive. 
Standard dark matter candidates include Weakly-Interacting Massive Particles 
(WIMPs), which have mass $\sim 0.1-1~\tev$, and arise from independent attempts 
in particle physics to understand the mechanism of electroweak symmetry breaking. 
Well-studied signatures of WIMPs include elastic scattering off nucleons in underground 
laboratories, missing energy signals at colliders, and particle production via self-annihilation 
in Galactic and extragalactic sources
(e.g.~\cite{Jungman:1995df,Baer:1997ai,Bergstrom:2000pn,Cheng:2002ej,Bertone:2004pz}). 

These standard signatures, however, do not provide the only means to test the 
properties of dark matter. As an additional example, one may also consider invoking 
dark matter physics to explain anomalous signatures in observed photon or cosmic ray 
emission spectra, even at energies seemingly far removed from those associated with 
weak-scale physics. In fact, in recent years, dark matter models have been constructed to 
explain astrophysical particle production over a variety of energy scales:  these include 
ultra-high energy cosmic rays~\cite{Berezinsky:1997hy}, 511 keV line emission from the 
Galactic bulge
~\cite{Boehm:2003bt,Picciotto:2004rp,Hooper:2004qf,Ferrer:2005xva,Kasuya:2006kj,Finkbeiner:2007kk,Pospelov:2007xh}, 
or the diffuse MeV gamma-ray background~\cite{Kribs:1996ac,Ahn:2005ck}. 
Though of course all of these anomalies will not be due to 
exotic dark matter physics, the observed emissions can 
provide strong constraints on well-motivated 
dark matter models, and further they may help identify the interesting 
regions of parameter space for a given dark matter model. 

In this paper we focus on two of the aforementioned anomalies: 
the diffuse MeV gamma-ray background and the 511 keV line flux from the
Galactic bulge. As we discuss below, known ``astrophysical" sources have both  
spectral shapes and rates that are unable to account for these observed emissions. 
The lack of well-motivated sources, as well as the similarities of energy scales, has given rise to 
speculation that both of these anomalies can be explained within the context of 
a single dark matter model~\cite{Lawson:2007kp}, albiet at the cost of abandoning 
the well-motivated WIMP framework. Remaining within the confines of WIMP models, 
these emissions have been separately reconciled by invoking unstable WIMPs with 
nearly degenerate, $\sim$ MeV scale mass splittings: the 511 keV line flux has
been studied in Refs.~\cite{Finkbeiner:2007kk,Pospelov:2007xh}, and 
the diffuse MeV photon background has been studied in Refs.~\cite{Cembranos:2007fj,Cembranos:2006gt}.  
The goal of this paper is to show that, remarkably, {\em both} the 511 keV line emission and the diffuse MeV gamma-ray 
spectrum can be explained in the context of a decaying WIMP model, characterized by 
a lifetime of $10^{20}$ s and near equal branching fraction to photons and electrons. 

Generically, we focus on a scenario in which the WIMP mass spectrum is 
highly degenerate, characterized by $\sim~\mev$ mass splittings between 
the next-to-lightest particle (NLP) and lightest particle (LP).  Both WIMPs freeze 
out under the standard conditions in the early Universe, and one of them is unstable 
with a lifetime in excess of the Hubble time. The NLP decays to three-body final states 
such as NLP 
$\rightarrow$ LP $+\gamma +\gamma$, NLP $\rightarrow$ LP $+ e^+ + e^-$, 
NLP $\rightarrow$ LP $+\bar{\nu} +\nu$. Two-body decays are assumed to be 
either highly suppressed or forbidden. Phenomenological consequences of 
two-body decays with similar lifetimes, and implications for the diffuse MeV photon background, were 
introduced in Ref.~\cite{Cembranos:2007fj}, and further discussed in Ref.~\cite{Yuksel:2007dr}. 
Additionally, prospects for detecting long-lifetime decaying dark matter with the GeV gamma-ray background 
were considered in Ref.~\cite{Bertone:2007aw}. 

As a specific implementation of the above idea, we consider the brane world 
scenario (BWS), which has become one of the most popular extensions to the 
Standard Model (SM). In the BWS, particles are confined to live on a three-dimensional 
brane embedded in a higher dimensional ($D=4+N$) space-time, while 
the gravitational interaction has access to the entire bulk space.
The fundamental scale of gravity in $D$ dimensions, $M_D$, can be lower than the 
Planck scale, $M_P$. In the original proposal~\cite{ArkaniHamed:1998rs,Antoniadis:1998ig}, 
the main aim was to address the hierarchy problem, and for that reason the value of $M_D$ 
was taken to be around the electroweak scale. However, brane 
cosmology models have also been proposed in which $M_D$ is much larger
than the TeV scale~\cite{Langlois:2002bb,Multamaki:2003zc}. In this paper, we consider a 
general BWS with arbitrary fundamental scale $M_D$; since we neglect
gravitational effects, our results do not depend on this scale once sufficiently high. 

In general, the existence of extra dimensions is responsible for the appearence of
new fields on the brane. On one hand, we have the tower of Kaluza-Klein (KK) 
modes of fields propagating in the bulk space, i.e. the gravitons. 
On the other, since the brane has a finite tension, $f^4$, its 
fluctuations will be parametrized by $\pi^\alpha$ fields 
called branons. When traslational invariance
in the bulk space is an exact symmetry, these fields can be understood as the massless
Goldstone bosons arising from the spontaneous breaking of that symmetry
induced by the presence of the brane~\cite{Sundrum:1998ns,Dobado:2000gr}.
However, in the most general case, translational invariance will
be explicitly broken and therefore we expect branons to be massive
fields. When branons are properly taken into account, the coupling of the SM
particles to any bulk field is exponentially suppressed by a
factor $\exp[-M^2_{KK}M_D^2/(8 \pi^2f^4)]$, where $M_{KK}$ is the
mass of the corresponding KK mode~ \cite{Bando:1999di,Cembranos:2005jc}. 
As a consequence, if $f\ll M_D$, the KK modes decouple from the SM particles. 
Therefore, for flexible enough branes, the only relevant degrees of freedom at
low energies in the BWS are the SM particles and branons.

The potential signatures of branons at colliders have been considered
in Refs.~\cite{Creminelli:2000gh,Alcaraz:2002iu,Cembranos:2003aw}, and 
astrophysical and cosmological implications have been studied in 
Refs.~\cite{Kugo:1999mf,Cembranos:2003aw}. 
Moreover in Ref.~\cite{Cembranos:2003mr} the 
possibility that massive branons could account for the observed dark matter
of the Universe was studied in detail (see also~\cite{Cembranos:2003aw}). 
Here, for the first time, we consider branon phenomenology with MeV mass
splittings and lifetimes in excess of the age of the Universe. Due to their 
universal coupling to the SM, decays of unstable branons produce electrons and photons at the
same rate in decays, leading to the aforementioned consequences for 
diffuse MeV gamma-rays and the 511 keV line flux. 

This paper is organized as follows. In section~\ref{sec:gammarayobservations}, we review 
the observations of the cosmic and Galactic gamma-ray backgrounds and the 511 keV signal. 
Section~\ref{sec:branons} we summarize the salient features of the brane world model, 
and in section~\ref{sec:signals} we present the resulting astrophysical signatures. 
Finally, in sections~\ref{sec:discussion} and~\ref{sec:conclusions} we discuss other possible 
signatures and recap our main conclusions. 

\section{Gamma-ray Observations}
\label{sec:gammarayobservations}

In this section, we introduce and discuss the gamma-ray emission spectra we will analyze 
within the framework of decaying dark matter. We focus on two specific observations: 
the 511 keV photon line flux from the Galactic center and
the high latitude isotropic diffuse MeV photon emission. We also, for completeness,
discuss the diffuse MeV gamma-ray emission from the Galactic center: 
in the discussion section below we show how these observations pertain to our model 
constraints. 

\subsection{511 keV line flux from the Galactic center} 

The SPI spectrometer on the INTEGRAL (International Gamma-ray Astrophysics Laboratory) 
satellite has measured a 511 keV line emission of $1.05 \pm 0.06 \times 10^{-3}$ photons 
cm$^{-2}$ s$^{-1}$ from the Galactic bulge~\cite{knodl2005}, confirming earlier measurements
~\cite{johnson1973}. The emission region is observed to be approximately spherically-symmetric 
about the Galactic bulge, with a full width half maximum (FWHM) $\sim 8^\circ$. There is a very low 
level detection from the Galactic disk, $\sim 4\sigma$, compared to the $50\sigma$ detection from 
the bulge. The 511 keV line flux is consistent with an $e^+ e^-$ annihilation spectrum, 
fit by the sum of three distinct components: a narrow and broad line flux, both centered at
511 keV, and a continuum spectrum extending to energies less than $511$ keV. 
The narrow line flux arises both from the direct annihilation of thermalized positrons 
into two photons or through para-positronium formation primarily in the cold and warm 
phase of the inter-stellar medium (ISM). 
The broad line flux, which has a width of $5.4 \pm 1.2$ keV FWHM, arises from the 
annihilation of para-positronium in flight in the warm and neutral phase of the ISM.  
The $< 511$ keV continuum emission arises from the annihilations in the 
ortho-positronium state.  The number of 511 keV gamma-rays produced 
per positron is $2(1-3p/4)$, where $p$ is the positronium fraction~\cite{Beacom:2005qv}. 
By detailed fitting of the spectrum to annihilation in different phases of the 
ISM, Ref.~\cite{Jean:2003ci} concludes that $p = 0.935_{-1.6}^{+0.3}$. A model independent fitting of 
the spectra to a broad line, narrow line, continuum spectra, and Galactic diffuse 
component concludes that $p = 0.967\pm0.022$. 
A precise determination of the positronium fraction thus ultimately 
depends on both the nature of the source of the positrons as well as the 
temperature of the medium in which they annihilate.  
 
The source of the Galactic positrons is uncertain. The fact that the annihilation
takes place primarily in the warm neutral and warm ionized medium implies that
the sources of the positrons are diffusely distributed, and that the initial kinetic energy
of the positrons is less than a few MeV~\cite{Beacom:2004pe,Beacom:2005qv}. The sources of the positrons 
are likely contained within the observed emission region; the 
propagation distance from creation to annihilation is at most of order $\sim 100$ pc.  
Type Ia supernovae (SNIa) are a candidate for the source of positrons, however 
recent estimates of the escape fraction of positrons from SNIa indicate that they cannot 
account for the entire 511 keV emission~\cite{Kalemci:2006bz}.
Several other astrophysical sources have been proposed~\cite{sturrock1971,ramaty1979,kozlovsky1987}, 
however none of these sources seem adequate to produce the intensity and spatial 
distribution of the observed line flux. 

\subsection{Isotropic diffuse MeV gamma-ray background at high Galactic latitude}
COMPTEL (the Compton Imaging Telescope) and SMM (the Solar Maximum Mission) 
have measured an isotropic gamma-ray background at high Galactic latitudes 
($\agt 10^\circ$) over the energy ranges $0.8-30~\mev$~\cite{Weidenspointner} and 
$0.3-7~\mev$~\cite{WatanabeSMM}, respectively. More recently,  INTEGRAL has measured 
a diffuse photon background over the energy range $5-100~\kev$~\cite{Churazov2007}. 
We follow the notation of Ref.~\cite{Yuksel:2007dr} and refer to these backgrounds as the isotropic 
diffuse photon background (iDPB), as it may include contributions from both Galactic and 
extragalactic sources. At $\sim$ MeV energies, the iDPB analysis is hindered by instrumental and 
cosmic-ray backgrounds, which must be carefully subtracted to reveal the underlying signal. 
Over the COMPTEL and SMM energy ranges, which will be important for our analysis 
below, the observed spectrum is observed to fall like a power law, with $dN/dE \sim E^{-2.4}$
~\cite{Weidenspointner}. 

Below energies of a few hundred keV, normal active galactic nuclei (AGN) 
are able to explain the mean flux of the cosmic X-ray background \cite{Ueda:2003yx}.   
A rare population of beamed AGN, or blazars, provide an important contribution
to the iDPB for energies $\agt 10$ MeV~\cite{Pavlidou:2002va},
though there is still room for other sources at these energies~\cite{LoebWaxman}. 
The iDPB is observed to smoothly transition in between these two energies, 
however, in the range $1~\mev \alt E_{\gamma} \alt
5~\mev$, no astrophysical source can account for the observed iDPB.
Blazars are observed to have a spectral cut-off $\sim 10~\mev$, and
also only a few objects have been detected below this
energy~\cite{McNaron-Brown}. SNIa contribute below $\sim 5~\mev$, but they also cannot account
for the entire spectrum~\cite{Strigari:2005hu,Ahn:2005ws}. Recent modeling
shows that nonthermal relativistic electrons which alter the AGN spectra 
for energies $\agt 1~\mev$ may account for this excess emission, though
a detailed understanding of the iDPB at all energies
$\agt~\mev$ will require matching the mean flux and the angular distribution
of the sources~\cite{Zhang:2004tj}. 

\subsection{Diffuse MeV gamma-rays from the Galactic center}   

COMPTEL has determined the flux spectrum of diffuse gamma-rays from the 
Galactic center region over the energy regime $1-20~\mev$ \cite{COMPTEL}. 
These fluxes have been averaged over a 
latitude of $|l| < 30^\circ$ and longitude $|b| < 5^\circ$, 
with high latitudes being used to define the zero flux level.  
It is currently unclear whether the $1-20~\mev$ spectrum is a result of diffuse or point 
source emission. 
For energies $\agt 100~\mev$, the gamma-ray spectrum is 
likely produced by both nucleon interactions with interstellar gas via neutral pion 
production and electrons via inverse compton scattering. For energies 
$\alt 100~\kev$, point sources dominate the gamma-ray spectrum~\cite{Lebrun2004}. 
However, an inverse compton spectrum that matches the diffuse gamma-ray spectrum
at $\sim 100~\mev$ can account for at most $50\%$ of the emission between 
$1-20~\mev$~\cite{Strong2000}.  

In addition to the diffuse measurement from COMPTEL, INTEGRAL has performed 
a search for gamma-ray lines originating within $13^\circ$ 
from the Galactic center. Two lines were recovered: the aforementioned line at 511 keV and 
an additional line at 1809 keV.  
The 1809 keV line originates from the hydrostatic nucleosynthesis of
radioactive elements in the cores of massive stars, where
long-lived isotopes such as $^{26}$Al are carried to the ISM, 
after which they subsequently decay mono-energetically. 
The 1809 keV line flux has been detected by INTEGRAL 
at a level of $\sim 10^{-4}$ cm$^{-2}$ s$^{-1}$. Other than these two gamma-ray 
lines, no other lines were detected up to upper flux limits of 
$10^{-5}$-$10^{-2}$ cm$^{-2}$ s$^{-1}$, depending on line width, energy, and exposure. 

\section{Branon Phenomenology}
\label{sec:branons}
Having introduced and discussed the gamma-ray emission spectra
we consider, 
in this section we will briefly review the main properties of massive
brane fluctuations (see Refs.~\cite{Dobado:2000gr,Cembranos:2001rp,Alcaraz:2002iu} 
for a more detailed description). Focusing in particular on models
that produce the observed abundance of dark matter, we present the formulae 
for determining the decay widths of long-lifetime branons to photons, electrons, 
and neutrinos. 

\subsection{Branon overview}

We consider a typical brane model in large extra dimensions. 
The four-dimensional space-time, $M_4$, is embedded in a
$D$-dimensional bulk space which, for simplicity, it is assumed to
factorize as $M_D=M_4\times B$. The extra space $B$ is a given
N-dimensional compact manifold, so that $D=4+N$. The brane lies
along $M_4$ and we neglect its contribution to the bulk
gravitational field. The bulk space
$M_D$ is endowed with a metric tensor $G_{MN}$, which
we will assume for simplicity is given by 
\begin{eqnarray}
 G_{MN}&=&
\left(
\begin{array}{cccc}
\tilde g_{\mu\nu}(x,y)&0\\ 0&-\tilde g'_{mn}(y)
\end{array}\right).
\end{eqnarray}
The position of the brane in the bulk can be parametrized as
$Y^M=(x^\mu, Y^m(x))$, with $M=0,\dots, 3+N$.  
We have chosen the bulk coordinates
so that the first four are identified with the space-time brane
coordinates $x^\mu$. We assume the brane to be created at a
certain point in $B$, $Y^m(x)=Y^m_0$, which corresponds to its
ground state. We will also assume that $B$ is a homogeneous space, so 
that brane fluctuations can be written in terms of properly normalized
coordinates in the extra space: $\pi^\alpha(x)=f^2 Y^\alpha(x)$, 
$\alpha=1,\dots, N$. The induced metric on the brane in its ground 
state is simply given by the
four-dimensional components of the bulk space metric, i.e.
$g_{\mu\nu}=\tilde g_{\mu\nu}=G_{\mu\nu}$. However, when brane
excitations  are present, the induced metric is given by
\begin{eqnarray}
g_{\mu\nu}=\partial_\mu Y^M\partial_\nu Y^N G_{MN}(x,Y(x)) =\tilde
g_{\mu\nu}(x,Y(x))-\partial_{\mu}Y^m\partial_{\nu}Y^n\tilde
g'_{mn}(Y(x))\,.
\label{induced}
\end{eqnarray}
The contribution of branons to the
induced metric  is then obtained by expanding Equation~\ref{induced} 
around the ground state \cite{Dobado:2000gr,Cembranos:2001rp,Alcaraz:2002iu}:
\begin{equation}
g_{\mu\nu}=
\tilde g_{\mu\nu}-\frac{1}{f^4}\delta_{\alpha\beta}\partial_{\mu}\pi^\alpha
\partial_{\nu}\pi^\beta
+\frac{1}{4f^4}\tilde g_{\mu\nu}M_{\alpha\beta}^2\pi^\alpha\pi^\beta
+\dots
\end{equation}
Branons are the mass eigenstates of the brane 
fluctuations in the extra-space directions. The branon mass matrix 
$M_{\alpha\beta}$ 
is determined by the metric properties of the bulk space and, 
in the absence of a general model for the
bulk dynamics, we will consider its elements as free parameters 
(for an explicit construction see Refs.~\cite{Andrianov:2003hx}). Therefore,
branons are massless only in highly symmetric cases
~\cite{Dobado:2000gr,Cembranos:2001rp,Alcaraz:2002iu,Cembranos:2004eb}.

Since branon fields survive in the limit in which gravity decouples, $M_D\rightarrow \infty$,
branon effects can be studied independent of gravity~\cite{Contino:2001nj}. 
We will work in the thin brane limit and assume that the 
brane dynamics can be described by a low-energy effective action
derived from the Nambu-Goto action
\cite{Dobado:2000gr}. 
 Also, 
branon couplings to the SM fields can be obtained from the SM action 
defined on a curved background given by the
induced metric Equation~\ref{induced}, and expanding in branon fields. 
Thus the complete action, up to second
order in $\pi$ fields, contains the SM terms, the kinetic term for the
branons and the interaction terms between the SM particles and the
branons \cite{Dobado:2000gr,Sundrum:1998ns,Cembranos:2001rp,Alcaraz:2002iu}.

It is interesting to note that under a parity transformation on
the brane, the branon field changes sign if the number of spatial
dimensions of the brane is odd, whereas it remains unchanged for even
dimensions. Accordingly, branons on a 3-brane are pseudoscalar
particles. This fact, in addition to the geometrical origin of the action, 
implies that terms in the effective Lagrangian with an odd number of 
branons are forbidden. It means that branons are stable. Strictly speaking,
this argument only ensures the stability of the lightest branon ($\bl$), 
the rest of branons can decay to the lightest one, producing SM particles. 
These decays are very suppressed if the branon spectrum is very degenerate. 

In order to analyze such properties we will study the simplest model containing 
an unstable branon. We consider a model where we have
two branons, implying that the number of extra dimensions has to 
be two or greater: the lightest one is defined as $\bl$, and the heaviest one as 
$\bh$. We will assume very  similar masses for them: 
$\Delta M\equiv M_{\bh}-M_{\bl} \ll M\simeq M_{\bh}\simeq M_{\bl}$. 
Using these energy eigenstates, the  SM-branon low-energy effective Lagrangian 
may be written as
\cite{Dobado:2000gr,Sundrum:1998ns,Cembranos:2001rp,Alcaraz:2002iu}:
\begin{eqnarray}
{\mathcal L}
_{\br-\phi}
&=&  \frac{I^\phi_{\alpha\beta}}{8f^4}(4\partial_{\mu}\br^\alpha
\partial_{\nu}\br^\beta-M^2\br^\alpha\br^\beta g_{\mu\nu})
T^{\mu\nu}_{\phi}
\,.\label{lag}
\end{eqnarray}
Here $\alpha(\beta)=h,l$, and $T^{\mu\nu}_{\phi}$ is the standard
energy-momentum tensor of the particle $\phi$ evaluated in the
background metric:

\begin{eqnarray}
T^{\mu\nu}_{\phi}=-\left(\tilde g^{\mu\nu}{\mathcal L}_{\phi}
+2\frac{\delta {\mathcal L}_{\phi}}{\delta \tilde
g_{\mu\nu}}\right)\,. 
\end{eqnarray}

The mass matrix is diagonal with eigenvalues $M_{\bh}\simeq M_{\bl}\simeq M$, whereas
the interaction matrix, $I_{\alpha\beta}$, is exactly the identity at first order 
\cite{Dobado:2000gr,Sundrum:1998ns,Cembranos:2001rp,Alcaraz:2002iu}. However, 
radiative corrections and higher order terms have a different effect on  
the interaction eigenstates  relative to the energy eigenstates, resulting in 
suppressed but non-zero cross interaction parameters, $\lambda_\phi$, 
\begin{eqnarray}
 I^\phi_{\alpha\beta}&\simeq&
\left(
\begin{array}{cccc}1&\lambda_\phi/2\\ \lambda_\phi/2&1
\end{array}\right).
\end{eqnarray}
This implies that, typically, $\lambda_\phi\sim 0.01$. For simplicity, we will take
this parameter to be independent of the particular SM particle, i.e. $\lambda_\phi\simeq \lambda$ 
for any $\phi$. The main conclusions of this paper do not depend on this assumption, 
although it does introduce an uncertainty of order one for a particular $\lambda_\phi$. 
However, for the most part of branon phenomenology, both this parameter 
and $\Delta M$ are negligible. General constraints on the branon parameter space we consider are 
shown in Figure~\ref{fig:CDM}: the constraints include those from HERA, Tevatron, and LEP-II 
(additional bounds from astrophysics and cosmology can be found in \cite{Cembranos:2003fu}).

Thus $\lambda$ is fundamental to analyze the stability of the heaviest branon. 
Indeed, if $\lambda$ is exactly zero, $\bh$ is completely stable.
If $\lambda$ is non zero, the heaviest branon decays to the lightest 
branon and a Standard Model particle anti-particle pair. For the analysis below, we are 
interested in $\Delta M<10\,\mev$, which means that the only available decay channels 
are $\Gamma_{e^+}:\bh\rightarrow\bl e^+e^-$, 
$\Gamma_{\gamma}:\bh\rightarrow\bl \gamma\gamma$, and $\Gamma_{\nu}:\bh\rightarrow\bl \nu\bar{\nu}$.

\subsection{Thermal abundances} 

Branons thus interact in pairs with the SM particles, and the lightest is necessarily stable. 
In addition, their couplings are suppressed by the brane tension 
$f^4$, which means that they could be in general weakly interacting and massive. 
As a consequence their freeze-out temperature
can be relatively high, which implies that their relic abundance can
be cosmologically significant. 

In order to calculate the thermal relic branon abundance, the standard  techniques have 
been used in the case of non-relativistic branons at decoupling
\cite{Cembranos:2003mr,Cembranos:2003fu}. The evolution of the number density 
$n_\alpha$ of branons interacting with SM particles in an 
expanding Universe is given by the Boltzmann equation:
\begin{eqnarray}
\frac{dn_\alpha}{dt}=-3Hn_\alpha-\langle \sigma_A v\rangle
(n_\alpha^2 -(n_\alpha^{eq})^2)\label{Boltzmann}
\end{eqnarray}
where
\begin{eqnarray}
\sigma_A=\sum_X \sigma(\pi^\alpha\pi^\alpha\rightarrow X)
\end{eqnarray}
is the total annihilation cross section of branons into SM particles $X$
summed over final states. The $-3Hn_\alpha$ term, with $H$ the Hubble parameter,
takes into account the dilution of the number density due to
the expansion of the Universe. This is the dominant mechanism that 
sets the total number density of branons. In the scenario we 
are interested in, the heaviest branon will also decay to the lightest branon,  
however this effect is insignificant in changing their respective number densities 
because we focus on lifetimes longer than the age of the Universe.
Therefore, the relic density is determined only by the parameters $M$ and $f$.  
Each branon species evolves independently and has exactly the same abundance 
before the decays are effective. 

\begin{figure}[bt]
\begin{center}
\resizebox{8.5cm}{!} {\includegraphics{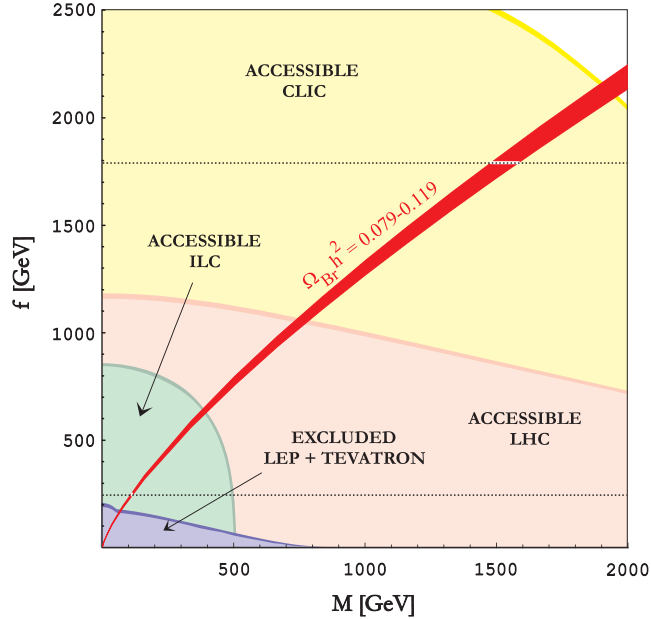}} 
\caption{The relic branon abundance for $N=2$ in the $f-M$
plane (see Refs.~\cite{Cembranos:2003fu,Cembranos:2003mr} for details). The
lower area is excluded by single-photon processes at LEP-II
\cite{Alcaraz:2002iu,Achard:2004uu}
together with monojet signals at Tevatron-I
\cite{Cembranos:2004jp}. Prospects for real branon production at
future colliders are also shown (See
Refs.~\cite{Alcaraz:2002iu,Cembranos:2004jp}). The region between the two
horizontal dotted lines are preferred by the muon anomalous magnetic
moment and electroweak precison observables
\cite{Cembranos:2005jc,Cembranos:2005sr}.
}
\label{fig:CDM}
\end{center}
\end{figure}

The relic abundance results for two branons of mass $M$ are shown in Figure~\ref{fig:CDM}.
As expected, for $f$ and $M$ scales of $\sim 100~\gev-1~\tev$, we have the 
correct amount of total non-baryonic dark matter abundance. 
As shown, these scales are not only in the 
natural region of parameter space, they are also in the favored region
for Brookhaven determinations of the muon anomalous magnetic moment
~\cite{Cembranos:2005jc,Cembranos:2005sr}.

\subsection{Branon decays}

As we have seen, decays of very degenerate branons can be 
described at low energies by an effective action that depends
on four parameters: the branon mass $M$, the mass splitting $\Delta M$,
the cross interaction parameter $\lambda$, and the
brane tension scale $f$, which suppresses the coupling of these 
new particles with the SM. We now determine the decay widths to 
the relevant SM particles. 

Since branons couple directly to the energy momentum tensor, 
they do not couple directly to a single photon, and thus decays into 
two-body final states that include a single photon are forbidden.
Indeed, this is a general property of scalar or pseudo-scalar particles; 
the decay of a spin-zero particle into another spin-zero particle and a photon is 
forbidden by angular momentum conservation. 
On similar grounds, the decay of a spin-zero particle into another spin-zero 
particle and a fermion is forbidden. Therefore, unless another light spin-zero particle 
is added to the Standard Model, the decays of degenerate scalar multiplets will 
proceed predominantly into three-body final states.

We focus specifically on the limit $M >> \Delta M$. In this limit, the 
decay widths can be calculated from Equation~\ref{lag}, the 
branon Feynman rules given in Ref.~\cite{Alcaraz:2002iu}, and 
substituting $\delta_{\alpha\beta}$ for $I_{\alpha\beta}$. For
photons, we get a differential decay width of 
\begin{equation}
\label{photon}
\frac{d\Gamma_{\gamma}(\varepsilon_\gamma)}{d\varepsilon_\gamma}=\frac{\lambda^2\,\varepsilon_\gamma^3\,
{\left(\Delta M-\varepsilon_\gamma   \right)}^3\,M^2}{24\,f^8\,{\pi }^3}, 
\end{equation}
which implies a total photon decay width of 
\hspace{-2cm}
\begin{equation}
\Gamma_{\gamma}= 
\left[1.67 \times 10^{20} s. 
\left[\frac{10^{-2}}{2\,\lambda}\right]^{2}
\left[\frac{f}{M}\right]^{2}
\left[\frac{4\, \mev}{\Delta M}\right]^{7}
\left[\frac{f}{1\, \tev}\right]^{6}
\right]^{-1}
\,.
\end{equation}
If $\Delta M > 2 m_e$, the heaviest branon can also decay to the lightest one
and an electron-positron pair.
The spectrum of the outgoing positron in the same limit of $M >> \Delta M $ is:  
\begin{eqnarray}
\label{positron}
&&\frac{d\Gamma_{e^+}(\varepsilon_{e^+},m_e)}{d\varepsilon_{e^+}}=\frac{\lambda^2\,M^2
}{2\,f^8\,{(4\pi)}^3}
\,\sqrt{(\varepsilon_{e^+}^2-m_e^2)((\Delta M-\varepsilon_{e^+})^2-m_e^2)}
\nonumber
\\
&&\,\,\,\,\,\,\,\,\,\,\,\,\,\,\,\,\,\,\,\,\,\,\,\,\,\,\,\,\,\,\,\,\,\,\,\,\,\,\,\,\,\,\,
\left\{
4(\Delta M-2\,\varepsilon_{e^+})^2[(\Delta M-\varepsilon_{e^+})\varepsilon_{e^+}
-16\,m_e^2]
\right.
\nonumber\\
&&\,\,\,\,\,\,\,\,\,\,\,\,\,\,\,\,\,\,\,\,\,\,\,\,\,\,\,\,\,\,\,\,\,\,\,\,\,\,\,\,\,\,\,\,\,
+\left.
m_e^2\,[225\,(\Delta M-\varepsilon_{e^+})\varepsilon_{e^+}-177\,m_e^2]
\right\}\,.
\end{eqnarray}
The kinematic limits for the total energy of the outgoing positron are $m_e>\varepsilon_{e^+}>\Delta M-m_e$.

\begin{figure*}[hbtp]
\begin{center}
\begin{tabular}{cc}
\includegraphics[height=6.5cm]{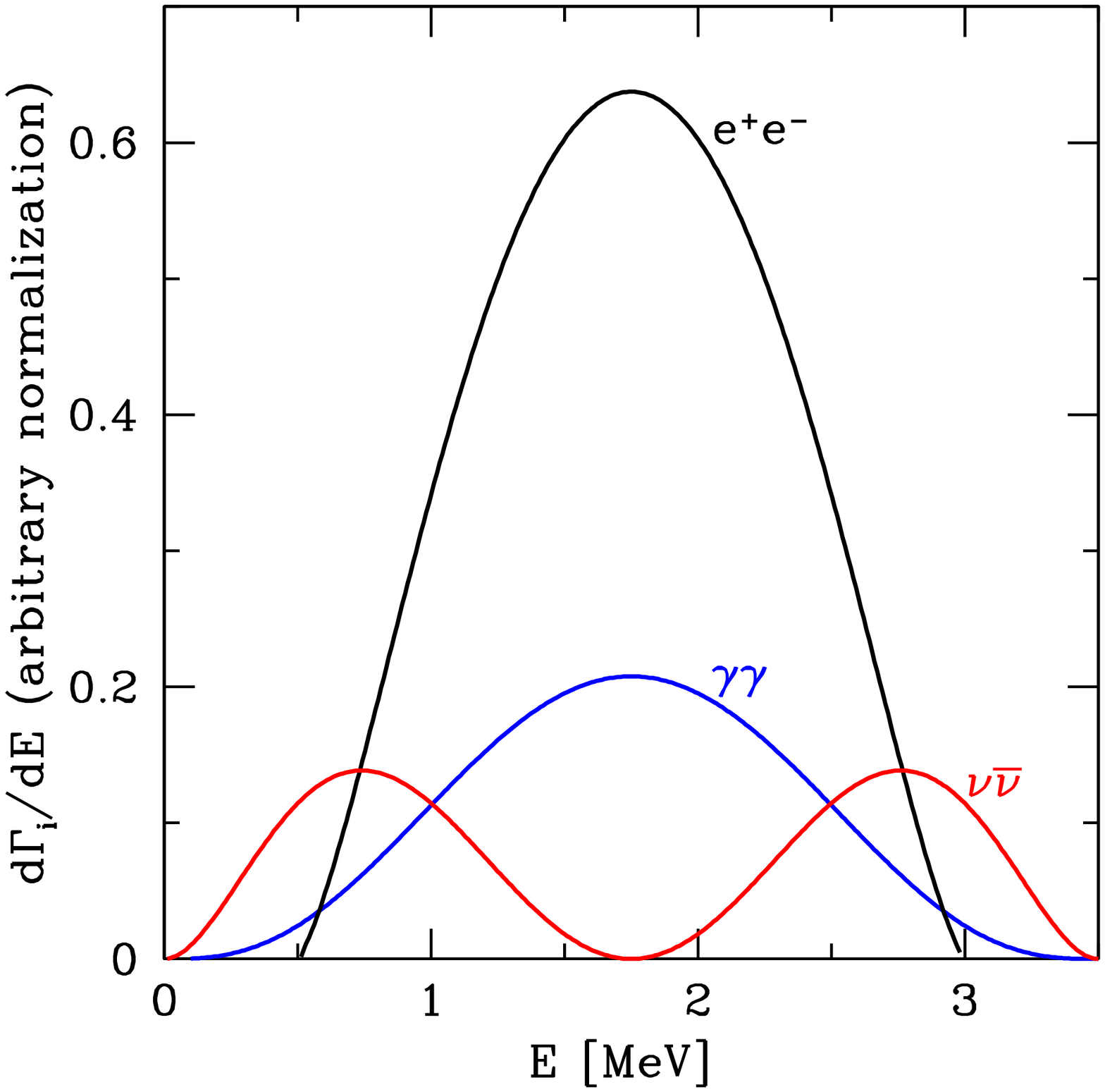}  &
\includegraphics[height=6.5cm]{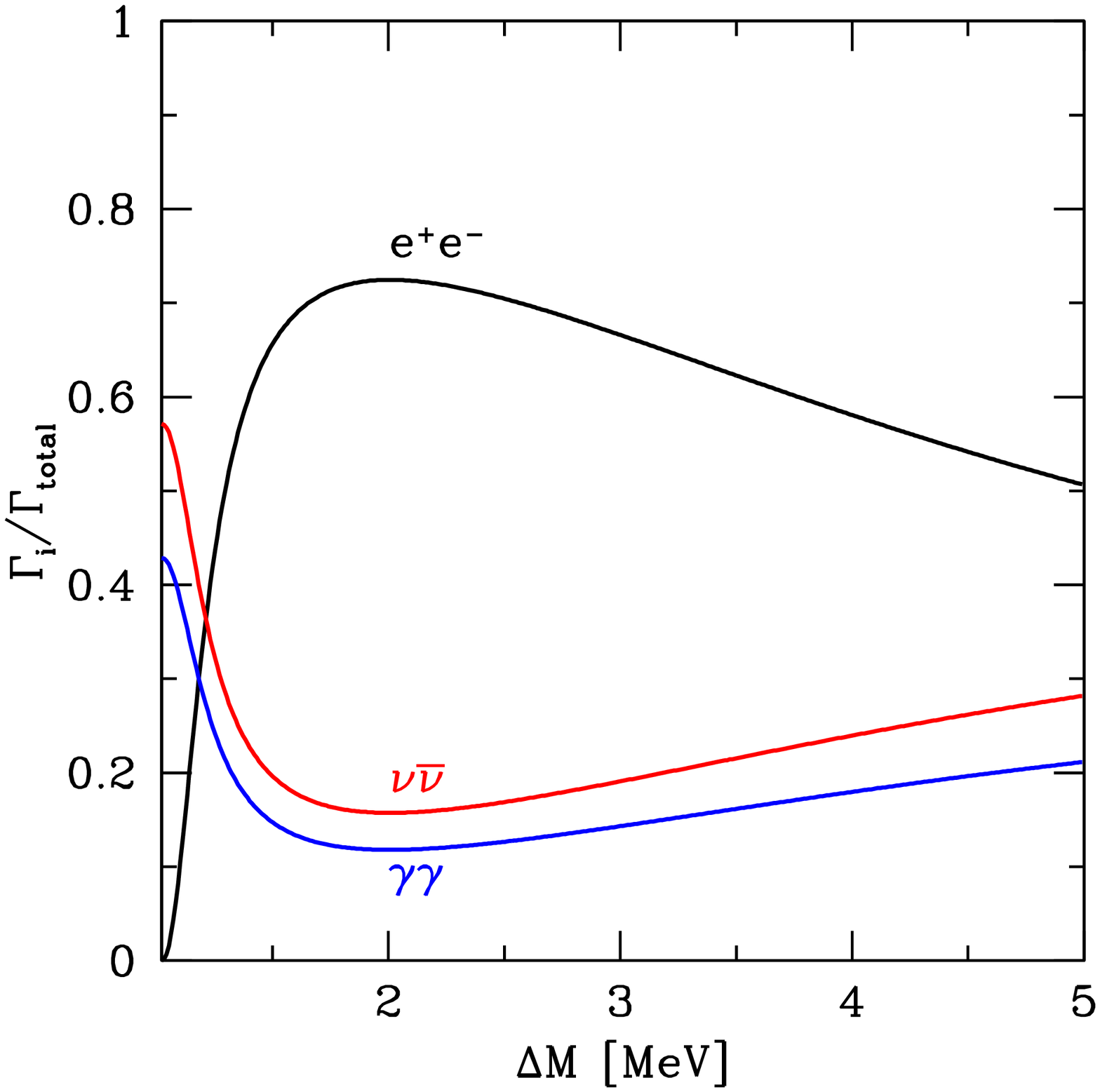} \\ 
\end{tabular}
\caption{\small {\em Left}: The differential decay widths for  $\bh \rightarrow \bl e^+e^-$, 
$\bl \gamma\gamma$, and $\bl \nu\bar{\nu}$, assuming $\Delta M = 3.5~\mev$. 
{\em Right}: The total branching fraction for each case as a function of $\Delta M$. 
In both figures, for $\nu\bar{\nu}$, we have accounted for the decays 
to all three species of neutrinos. 
\label{fig:branching}
}
\end{center}
\end{figure*}

Finally, we note that in the kinematic limits we study it is also possible to 
produce neutrinos: 
\begin{eqnarray}
\label{neutrino}
\frac{d\Gamma_{\nu_i}(\varepsilon_{\nu_i})}{d\varepsilon_{\nu_i}}
&=&\frac{1}{2}
\frac{d\Gamma_{e^+}(\varepsilon_{\nu_i},0)}{d\varepsilon_{\nu_i}}\,,
\end{eqnarray}
where $i$ labels the species of neutrino (we have neglected the neutrino mass). 
Since diffuse neutrino bounds are much weaker than diffuse photon bounds at MeV 
energies, there are likely minimal phenomenological implications of neutrino production
in this model (see Refs.~\cite{Yuksel:2007ac,PalomaresRuiz:2007ry} 
 for recent determinations of neutrino constraints of annihilating and decaying dark 
matter models). The decay widths for all of these channels are shown in Figure~\ref{fig:branching}.

\section{Galactic and Extragalactic Signals} 
\label{sec:signals}

We now determine the extragalactic and Galactic gamma-ray signals from the decays of
unstable $\bh$'s. Although the resulting fluxes of course depend on the specific model 
parameters and couplings, the formalism we present here can be used for any similar 
model that decays via three-body final states. In what follows, the mass of the NLP is defined
as $M$, the mass splitting between the NLP and LP is $\Delta M$, and the lifetime of the 
NLP into a given branching ratio, $\imath$, as $1/\tau^\imath = \Gamma_\imath = 
\int dE d\Gamma_\imath/dE$. 

\subsection{Galactic 511 keV flux}
We begin by considering the 511 keV line emission resulting from positron injection in the 
Galactic bulge. Given the small propagation distance of the positrons as described
above, we assume that the location of the $e^+ e^-$ annihilation and 511 keV photon 
creation faithfully tracks the dark matter halo distribution. 
Defining $s$ as the distance from the 
Sun to any point in the halo, $\Psi$ as the angle between the Galactic center and any point
in the halo, and $D=8.5$ kpc as the distance from the Sun to the Galactic center, the 511 keV line 
intensity from decays is 
\begin{equation}
\frac{d\Phi_{511}(\Psi)}{d\Omega} = 
\frac{1}{4 \pi}\int_0^\infty \frac{d n_{511}[r(s,\Psi)]}{dt} ds\,,
\label{eq:GalacticFlux}
\end{equation}
where $r^2(s,\Psi)= D^2 + s^2 - 2Ds\cos{\Psi}$, and the solid angle is 
defined as $\Delta \Omega = 2\pi(1-\cos \Psi)$. 
The number density of dark matter particles is $\rho(r)/M$, where 
$\rho(r)$ is the density of dark matter in the halo
of the Milky Way, which we assume to be spherically symmetric. 
Consequently, the total number of 511 keV photons produced 
per unit time can be estimated as $d n_{511}/dt= 2(1-3p/4)n_{\bh} \Gamma_{e^+}=
(1-3p/4)\,\rho\,\Gamma_{e^+}/M$, where we are assuming that $\bh$ accounts for half of the dark matter 
(the other half is in form of $\bl$), that the positrons are stopped before annihilation, and that 
this annihilation takes place through positronium formation $p$ fraction of the time. As 
we have discussed above, $p\simeq 0.94$.

Both the total flux within the field-of-view and the angular distribution of the flux
depend crucially on the shape of the dark matter density profile. 
Numerical simulations with only dark matter have shown that the halos consisting
of cold dark matter particles have a Navarro-Frenk-White (NFW) type density profile, 
$\rho(r) = \rho_0 /(r/r_0)/(1+r/r_0)^2$, where $\rho_0$ is the scale density and 
$r_0$ is the scale radius \cite{Navarro:1996gj}. The NFW profile, however, does not
account for 
the effect of baryonic physics, which is expected to be important in setting the central 
density of Milky Way-type galaxies,  
due to the relatively large baryonic contribution in the central regions. 
A full understanding of the halo density profile must adequately account for the energy 
exchange processes between the baryons and dark matter, which is currently lacking. These interactions are particularly 
important in the interior regions, and it is currently not understood if these interactions 
steepen the central density of the dark matter halo or flatten it out. 
From an observational perspective, there is also a wide uncertainty in the density profile 
and mass model of the Milky Way halo. For example, Ref.~\cite{Binney:2001wu} argues that the large number 
of microlensing constraints towards the Galactic bulge is inconsistent with central slopes 
steeper than $r^{-0.4}$. Ref.~\cite{Klypin:2001xu} apply a variety of observational constraints to an adiabatically- 
contracted NFW model, and find that NFW-like central slopes provide an excellent fit to 
the data. 

To allow for the greatest flexibility, we model the density profile of the Milky Way in the form of
\begin{equation}
\rho(r)= \frac{\rho_0 }{{(r/r_0)}^{\gamma}
\,{\left[1+{(r/r_0)}^\alpha\right]}^{(\beta-\gamma)/\alpha}
}, 
\label{eq:densityprofile}
\end{equation}
where $\rho_0$ is the scale radius, $r_0$ is the scale density. The combination of
$\gamma$ and $\beta$ set the inner and outer slopes of the halo profile, respectively, 
while $\alpha$ controls the sharpness of the transition between the inner and outer slopes. 
We determine the shape parameters $\alpha$, $\beta$, and $\gamma$, as well as the normalization
parameters $\rho_0$ and $r_0$ by fitting to both the angular distribution of the INTEGRAL signal and 
the constraints on the Milky Way halo. The parameters $\rho_0$ and $\gamma$ are primarily 
determined by the INTEGRAL data, whereas the remaining parameters are fixed by the total mass
and local density of the dark matter halo. 
The intensity of the INTEGRAL signal is so low at high
longitudes that it is difficult to discriminate from the instrumental background, and thus fix the parameters
that do not directly effect the central region of the halo.  

\begin{figure}
\resizebox{7.5cm}{!}{
\includegraphics{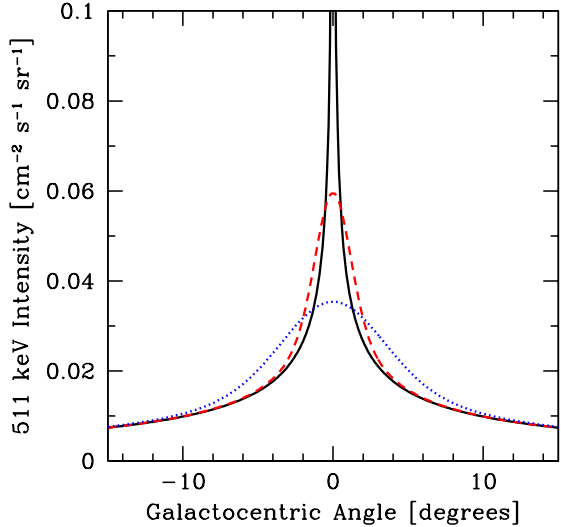}
} 
\caption{The 511 keV intensity from the decays of unstable $\bh$'s as a function of Galactocentric angle. 
The intensity has been normalized
to the total flux of $1.05 \times 10^{-3}$ cm$^{-2}$ s$^{-1}$ within the Galactic bulge, 
and the dark matter halo is described by Equation~\ref{eq:densityprofile} with $\alpha = 8$, $\beta = 3$, and
$\gamma=1.5$.  
The solid (black) line is the true signal, the dashed (red) line is the signal 
smoothed with a $1^\circ$ angular resolution, and the dotted (blue) curve is 
for a $3^\circ$ angular resolution. We have used a positronium fraction $p=0.94$, 
and the branon model parameters $\Delta M = 3.5~\mev$, $f = 1~\tev$, and
$M_{\bh} = 700~\gev$. The angular distribution of the INTEGRAL signal is spherically- 
symmetric about the Galactic bulge, with a FWHM of $8^\circ$. 
\vspace*{-.2in}
\label{fig:511}}
\end{figure}

In Figure~\ref{fig:511}, we show an example halo profile that fits the angular distribution
of the 511 keV intensity, and is normalized to match the total flux of $1.05 \times 10^{-3}$ cm$^{-2}$ s$^{-1}$ 
within the Galactic bulge. For simplicity, we assume that the 511 keV emission is entirely a result of 
dark matter decays, and does not include any contribution from sources discussed above, such as SNIa. Here we use  
$\Delta M= 3.5~\mev$, $M_{\bh} = 700~\gev$, $\lambda = 0.0085$, and a brane tension of $f = 1$ TeV;
these model parameters naturally give the correct relic dark matter abundance, as seen in 
Figure~\ref{fig:CDM}. The implied parameters describing the halo model are $\rho_0= 0.12$ GeV cm$^{-3}$, 
$r_0= 10$ kpc, $\gamma=1.5$, $\beta=3$, $\alpha=8$. 
With these halo model parameters, the total dark matter halo mass is
in accord, to within statistical uncertainties, with the mass
determinations from the kinematics of satellite galaxies~\cite{Wilkinson:1999hf,Sakamoto:2002zr}.
In addition to the true signal, which is shown as 
a solid line in~Figure~\ref{fig:511}, we show two curves with angular resolutions of 
$1^\circ$ and $3^\circ$; these are similar to the $\sim 3^\circ$ FWHM angular resolution of SPI. 

In Figure~\ref{fig:511}, we have focused on matching the normalization and angular 
distribution of the flux within the Galactic center region, where the 511 keV flux
is well-determined. As an additional check on the viability of the model, we must be 
sure not to overproduce the observed 511 keV emission from high longitudes in the 
Galactic disk. We can examine constraints on a potential high longitude disk flux by separately considering 
both emission from the Galactic center region and large-scale extended diffuse emission. Regarding the 
Galactic center region emission, we use the most recent INTEGRAL results,  which shows that the 511 keV 
bulge-to-disk flux ratio is $\sim 1-3$ and that the intensity is reduced by a factor $\sim 6$ at latitudes and
longitudes of $10^\circ$ relative to the Galactic center~\cite{knodl2005}. As seen in~Figure~\ref{fig:511},
the chosen profile satisfies these constraints. Regarding the large scale ($\agt 20^\circ$)
diffuse 511 keV emission, we consider the results of Ref.~\cite{Teegarden:2006ni}.  
Using the INTEGRAL effective area of 75 cm$^{2}$ at 511 keV, and smoothing the 
solid curve in Figure~\ref{fig:511} with a gaussian comparable to
the $\sim 20^\circ$ FWHM spatial scales resolved in Ref.~\cite{Teegarden:2006ni}, 
we find a 511 keV rate of $\sim 0.06$ s$^{-1}$ at $l = 0^\circ$, and $\sim 0.01$ s$^{-1}$ at 
$l = 30^\circ$. These rates were determined by subtracting the flux at $b = 20^\circ$, and are 
consistent with the rates determined in Ref.~\cite{Teegarden:2006ni}. Thus we find that the halo 
profile used in Figure~\ref{fig:511} is consistent with 511 keV emission constraints at high Galactic
longitude. 

Before moving on to consider the direct production of gamma-rays in $\bh$ decays, we 
make one additional comment regarding the assumed value of the positronium fraction.  
Motivated by the detailed fits by INTEGRAL, for all of the above results we have taken the positronium 
fraction to be $p=0.94$. However, the precise determination of $p$ is model dependent, as 
discussed above. If the decays we consider do indeed  
account for a large fraction of the 511 keV signal, a redetermination of $p$ will be required 
by explicitly fitting both the Galactic and extragalactic fluxes. Inclusion of the latter flux will
be important, even in the direction of the Galactic center, given that the extragalactic and Galactic 
fluxes are similar to within an order of magnitude, depending specifically on the shape of the
halo model. The similarity of extragalactic and Galactic fluxes is unique to the models we consider; 
in the case of dark matter annihilation, for example, the extragalactic flux is typically several 
orders of magnitude below the flux from any direction in the Galaxy itself. 

\subsection{High latitude gamma-ray flux} 

We now turn to determining the contribution to the high latitude iDPB from three-body decays to 
gamma-rays, both from the Galactic halo and from extragalactic sources. First considering
extragalactic sources, these decays will be visible only if they occur in the late Universe, 
in the matter or vacuum dominated eras.  Assuming a smooth distribution of dark 
matter in the Universe (modifications to this assumption will be discussed below), 
the differential gamma-ray flux from a general three-body decay is
\begin{equation}
\frac{d\Phi}{ dE_{\gamma}}
=\frac{n_\gamma}{4\pi} \int_0^{t_0} dt
\frac{N(t)}{a\,V_0}
\frac{d\Gamma_{\gamma}}{d\varepsilon_\gamma}\,.
\label{eq:extragalactic}
\end{equation}
Here $n_\gamma$ is the number of photons produced in a single decay;
$a$ is the scale factor of the Universe; $t_0 \simeq 4.3\times 10^{17}$ s is 
the age of the Universe; $N(t) = N^{\text{in}} e^{-t/\tau}$, where $N^{\text{in}}$ 
is the number of decaying particles at freeze-out; and $V_0$ is the present 
volume of the Universe. The relation between the produced energy, $\varepsilon_\gamma$,
and observed energy, $E_{\gamma}$, is given by 
$d\varepsilon_\gamma/dE_{\gamma}=(1+z)\equiv a^{-1}$. In Equation~\ref{eq:extragalactic},
we have assumed that gamma-rays do not get attenuated on cosmological gamma-ray
backgrounds from their redshift of production. We find this to be an excellent approximation
over the energies and redshifts we are considering \cite{Chen:2003gz}. In the BWS,  
$n_\gamma = 2$, however this factor gets cancelled because $1/2$ of the dark matter
consists of unstable $\bh$'s. 

We are interested in decays such that $\tau \gg t_0$. In this case, we can 
write $t\simeq P(a)$, where 
\begin{equation}
P(a)\equiv \frac{2\left(
\ln{\left[\sqrt{\Omega_{\Lambda}a^3}
+\sqrt{\Omega_{M}+\Omega_{\Lambda}a^3}\right]}
-\ln{\left[\sqrt{\Omega_{M}}\right]}
\right)}{3 H_0 \sqrt{\Omega_{\Lambda}}}
\,. 
\label{eq:Pequation}
\end{equation}
The differential photon flux coming from the decays may be written as
\begin{equation}
\frac{d\Phi}{ dE_{\gamma}} \simeq
\frac{n_\gamma}{4\pi}\frac{N^{\text{in}}}{V_0}\int_{E_{\gamma}/\Delta M}^{1} da
 \frac{e^{-P(a)/\tau}}{aQ(a)}
 \left.\frac{d\Gamma_{\gamma}}{d\varepsilon_\gamma}
 \right|_{\varepsilon_\gamma=E_\gamma/a}, 
\label{eq:phi}
\end{equation}
where we have defined 
\begin{equation}
\frac{da}{dt}= a H
\simeq H_0\sqrt{
\frac{\Omega_{M}}{a}
+\Omega_{\Lambda}a^2} 
\equiv Q(a).  
\end{equation} 
For lifetimes much longer than the age of the Universe, the exponential factor
in Equation~\ref{eq:phi} is negligible. We neglect the radiation content,
$\Omega_{R}\sim 0$, as well as the curvature term, $k\sim 0$.
Though the extragalactic flux is dependent on cosmological parameters~\cite{Cembranos:2007fj}, 
here we choose to fix the parameters to match the observed concordance cosmological model~\cite{Spergel:2006hy}. 

\begin{figure}
\resizebox{3.25in}{!}{
\includegraphics{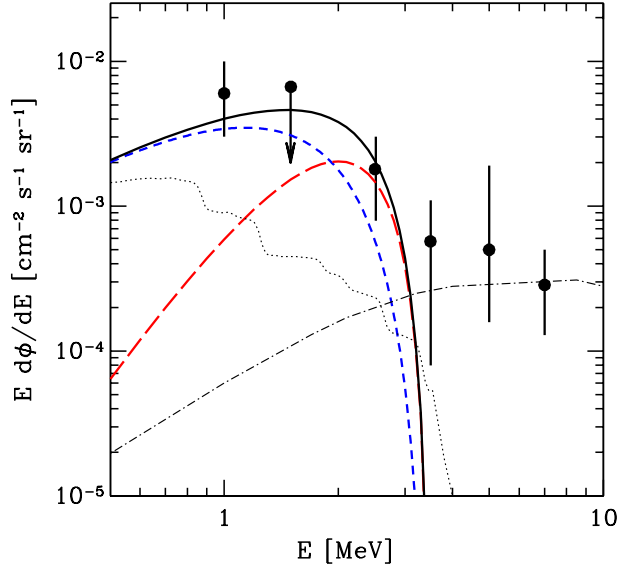}
} 
\caption{The high latitude isotropic diffuse photon background in the MeV regime, as measured by 
COMPTEL (circles)~\cite{Weidenspointner}. The short-dashed curve is the contribution from extragalactic
decays, the long-dashed curve is the contribution from decays in the Galactic halo, and the solid curve
is the sum of both extragalactic and Galactic contributions. The branon model parameters are the same as those
in Figure~\ref{fig:511}.  Two additional standard contributions to the iDPB are shown: Type Ia
Supernovae (SNIa; dotted) and blazars (dot-dashed). The blazar background has been 
normalized to the gamma-ray background at energies $\agt~10~\mev$~\cite{McNaron-Brown}, 
while the normalization of the SNIa spectrum is taken from Ref.~\cite{Strigari:2005hu}. 
\vspace*{-.2in}
\label{fig:idpb}}
\end{figure}

The high latitude iDPB also receives a significant contribution from decays in the halo 
of the Galaxy. The flux from Galactic halo dark matter decays directly into photons can also be calculated using
Equation~\ref{eq:GalacticFlux}, accounting for the fact that the flux comes from high 
latitudes, $\agt 10^\circ$, and averaging the flux over a $\sim 10^{-3}$ sr field of view.  
This latter quantity corresponds to the field of view characteristic of MeV gamma-ray instruments 
such as COMPTEL. 
With BWS parameters fixed by the 511 keV signal, in Figure~\ref{fig:idpb}
we show the contribution to the iDPB from high latitude Galactic and extragalactic $\bh$ 
decays. Again, unstable $\bh$'s account for $1/2$ of the relic abundance of dark matter. 
In the calculation of the observed spectrum, 
we have not accounted for finite detector energy resolution; including these effects 
will likely further smooth out the endpoints of the spectrum, and bring the shape of
the spectrum into even better agreement with the approximate power-law spectrum observed
by COMPTEL. 

\section{Discussion}
\label{sec:discussion} 

Above we have shown that, assuming both an abundance of non-baryonic dark matter
that matches cosmological observations and WIMP scale masses, within the BWS 
two parameters are required to independently match the 511 keV line flux and the 
$\sim~\mev$ iDPB. We now discuss independent constraints on this model as
well as potential implications of these results. 

\subsection{Gamma-rays from the Galactic center} 
In section~\ref{sec:signals} we have focused on the high latitude Galactic gamma-ray 
background. There will additionally of course be a significant contribution to the low-latitude 
gamma-ray background from $\bh$ decays, in particular in the direction the Galactic center.  
We must be sure that the model parameters we consider above do not violate the COMPTEL
flux measurements in this direction. 

We can make a simple estimate of the gamma-ray flux in the direction of the Galactic 
center by considering the branching ratios determined in section~\ref{sec:branons}. 
Using these branching ratios, in combination with the number of gamma-rays produced per $e^+ e^-$ annihilation
(section~\ref{sec:gammarayobservations}), the ratio of the total number of photons produced at 511 keV 
to the total number of photons produced directly in the two photon decays for our 
model parameters is simply 
\begin{equation}
\frac{\Gamma_e}{\Gamma_\gamma} \left(1-\frac{3}{4}p \right) \simeq 0.9. 
\label{eq:gamma511ratio}
\end{equation}
In general, we can expect a ratio of order one for any model that has similar branching
ratios to photons and positrons. Thus we expect a continuous gamma-ray flux    
$\alt 10^{-2}$ cm$^{-2}$ s$^{-1}$ sr$^{-1}$ at $\sim$ MeV energies from the Galactic
center. Interestingly, COMPTEL has measured a diffuse flux from the Galactic center 
at this order of magnitude in the energy range $1-3~\mev$~\cite{Strong2000}. The COMPTEL flux is seen 
to vary slowly with energy, and is reduced by a factor $\sim 2$ at a Galactic latitude 
of $\sim 10^\circ$. We find that, for the dark matter halo profile considered above, the 
diffuse MeV gamma-ray emission from $\bh$ decays is consistent with the COMPTEL flux 
both at the Galactic center and at higher latitudes. Since, as discussed above, the source of MeV 
gamma-rays from the Galactic center is uncertain, it is interesting to consider 
dark matter decays as the source of this emission. 

In addition to the diffuse Galactic gamma-rays, INTEGRAL has placed 
strong constraints on the presence of anomalous gamma-ray lines in the
energy regimes $\sim 10-8000~\kev$ within $13^\circ$
from the Galactic center~\cite{Teegarden:2006ni}. For example, at energies $\sim~\mev$, 
lines of width $\sim~\kev$ are constrained to have a flux $\alt 10^{-4}$ cm$^{-2}$ s$^{-1}$,
and ``lines" of width $\sim$ MeV are constrained to have a flux $\alt 10^{-3}$ cm$^{-2}$ s$^{-1}$. 
Given the width of the gamma-ray spectra we consider (see Figure~\ref{fig:branching}), the latter 
flux limits are relevant for $\bh$ decays.  
As a result, the INTEGRAL line constraints are no more stringent than the COMPTEL constraints
on diffuse MeV gamma-ray emission. 

\subsection{Assumption of smooth halo and angular distribution}

The extragalactic flux from dark matter decays depends on the distribution of
dark matter halos in the Universe, while the Galactic flux depends on the 
distribution of dark matter within the Milky Way halo. In principle, both of these
fluxes also depend on the presence of ``substructure" within dark matter halos
~\cite{Diemand:2006ik}. 
While we have assumed that the distribution of dark matter is smooth both 
in the Universe and in the Galactic halo, we can gain some insight as to the 
effects of clumping by considering the simplifying limit in 
which all of the substructures are at a common mass, the internal density of the substructures is constant, and 
the radial distribution of substructures in the halo follows the same density profile as the smooth 
dark matter halo. In reality, it is likely that the two radial distributions differ, which may 
affect our simple estimate. Under these assumptions, we find that, in a given direction, the 
increase in the flux over the smooth halo is simply $1+f$, where $f$ is the fraction of
the halo contained in substructures. Typically, $f \simeq 0.01$~\cite{Diemand:2006ik}. 
This small increase simply quantifies the probability that any one of these substructures 
is within a given field of view. 

Regarding the determination of the extragalactic gamma-ray flux from $\bh$ decays, 
we have assumed a smooth distribution of $\bh$'s in the Universe, and have neglected
the clumping in dark matter halos. Given the above considerations, the clumping in halos 
will likely have minimal effect on the mean flux signal, though the situation will be different when 
considering the angular distribution of gamma-rays. The angular distribution depends not
only on the parameters describing the branon model, but also on the cosmological 
evolution of dark matter halos as a function of redshift. Similar to the case of standard
annihilating neutralino WIMP dark matter, the angular distribution of gamma-rays from 
decays is expected to be distinct from astrophysical sources such as AGN and supernovae, 
and may provide a way to discriminate between the sources of the iDPB~\cite{Ullio:2002pj,Ando:2006cr}. 
We consider this analysis in a forthcoming study. 

\section{Conclusions}
\label{sec:conclusions} 

In this paper, we have analyzed the general possibility that decaying dark matter may
be the source of both the 511 keV gamma-ray line from the Galactic center 
and the diffuse high latitude photon background at energies between 1 and 5 MeV. 
We have introduced a new scenario where a dark matter particle is nearly degenerate
with a lighter daughter, with mass splitting $\sim$ MeV and lifetime $\sim 10^3$
times the present age of the Universe. The decays of the dark matter are into three
body final states: a Standard Model particle anti-particle pair and a daughter dark matter particle. 
Remarkably, we find that the above mass splitting and lifetime are exactly that which is 
required to explain both of the above gamma-ray emissions, provided there is similar
branching ratios to electrons and photons. 

We have illustrated this idea with a concrete model of brane-world dark matter, and
we have shown that a standard WIMP motivated within the brane-world scenario, the
branon, can explain both observations with natural values required to obtain the correct
dark matter relic abundance. Using the effective low-energy Lagrangian for a model with 
two branons interacting with the Standard Model fields, we have computed the decay rates 
of the heaviest branon into the lighter one and pairs of Standard Model particles, in particular 
focusing on electron-positron pairs and photons. We find that the branon model not only produces
positrons and gamma-rays at the required rates, it also provides unique and novel 
phenomenology that will be tested by future MeV gamma-ray observatories.  

\begin{acknowledgments}
We are very grateful to Jonathan Feng, Matt Kistler, and especially John Beacom and 
Hasan Yuksel for several useful discussions on this paper. The work of JARC is supported 
in part by DOE grant DOE/DE-FG02-94ER40823,
by the FPA 2005-02327 project (DGICYT, Spain) and the
McCue Award from the UCI Center for Cosmology. LES acknowledges 
support from NSF grant AST-0607746. 
\end{acknowledgments}

\bibliography{branon}
	      
\end{document}